\documentclass[aps,pre,preprint,groupedaddress,letterpaper]{revtex4}
\usepackage[dvips]{graphicx}       
\DeclareGraphicsExtensions{.eps}   

\graphicspath{{./figs/}}
\usepackage{bm}\def\vec#1{\bm{#1}}

\begin{document}

\title{A Biomechanical Model for Dictyostelium Motility}

\author{Mathias Buenemann}
\affiliation{%
Center for Theoretical Biological Physics,
University of California, San Diego, La Jolla, CA 92093-0374
}%
\author{Herbert Levine}
\affiliation{%
Center for Theoretical Biological Physics,
University of California, San Diego, La Jolla, CA 92093-0374
}%
\author{Wouter-Jan Rappel}
\affiliation{%
Center for Theoretical Biological Physics,
University of California, San Diego, La Jolla, CA 92093-0374
}%
\author{Leonard M. Sander}
\affiliation{%
Department of Physics and Michigan Center for Theoretical Physics,
University of Michigan, Ann Arbor, Michigan 48109, USA
}%

\begin{abstract}
  The crawling motion of \textsl{Dictyostelium discoideum} on
  substrata involves a number of coordinated events including cell
  contractions and cell protrusions.  The mechanical forces exerted on
  the substratum during these contractions have recently been
  quantified using traction force experiments.  Based on the results
  from these experiments, we present a biomechanical model of
  \textsl{Dictyostelium discoideum} motility with an emphasis on the
  adhesive properties of the cell-substratum contact.  Our model
  assumes that the cell contracts at a constant rate and is bound to
  the substratum by adhesive bridges which are modeled as elastic
  springs.  These bridges are established at a spatially uniform rate
  while detachment occurs at a spatially varying, load-dependent rate.
  Using Monte-Carlo simulations and assuming a rigid substratum, we
  find that the cell speed depends only weakly on the adhesive
  properties of the cell-substratum, in agreement with experimental
  data.  Varying the parameters that control the adhesive and
  contractile properties of the cell we are able to make testable
  predictions.  We also extend our model to include a flexible
  substrate and show that our model is able to produce substratum
  deformations and force patterns that are quantitatively and
  qualitatively in agreement with experimental data.
\end{abstract}
\maketitle 

\section{Introduction}

Cell movement over solid surfaces plays a key role in many every-day biological processes
including embryogenesis, osteogenesis, wound healing, and immune defense
\cite{franz02}. For example, neutrophils 
chemotax towards a wound in order to prevent infection~\cite{baggiolini98}. 
On the other hand, cell motility can play a significant role in disease; for instance, cancer cells spread out and intrude
into healthy tissue by directed, active motion ~\cite{wang05,condeelis04,kedrin07}.  Hence, deeper
insight into the biochemical and mechanical processes involved in cell
crawling would be of great interest and importance. 

Despite their apparent differences, many eukaryotic cells share
essential characteristics of their crawling motion~\cite{rafelski04,mogilner09}.  At the macroscopic level, cell motion often consists of several  distinguishable phases: (i) extension of a membrane protrusion
(pseudopod) at the leading edge, (ii) attachment of the pseudopod to
the substratum, and (iii) detachment and subsequent retraction of the
cell rear. These mechanical changes are mainly
driven by polymerizing F-actin (protrusion) and myosin motors
(retraction) \cite{mogilner09}. 
Both processes are regulated and synchronized in a
spatio-temporal manner~\cite{lauffenburger96}.
Additionally, in many higher organisms, detachment is regulated via
biochemical changes of focal adhesions
\cite{lee97,laukaitis01,kaverina02}.  
In other motile cells, on the other hand,
focal adhesions are absent and
a similar degradation mechanism has not yet been reported.

Much of our understanding of cell motility has come from
experiments on the social amoeba \textit{Dictyostelium discoideum}
which has been established as an experimental model system
during the past decades~\cite{parent99,noegel00,kessin01}.  These cells move 
rapidly ($\sim$ 10 $\mu $m/min) 
and can be very sensitive to chemical cues. Also, the availability
of a large variety of mutants allows quantitative insight into
regulatory as well as mechanical aspects of cell motion. This paper is devoted to presenting a simple model for \textit{Dictyostelium} crawling, with specific emphasis on the biomechanics of adhesive contacts between the cells and the 
substratum. 
  
One motivation for this study relates to recent force cytometry experiments in which the traction forces exerted by
motile  \textit{Dictyostelium}  cells chemotaxing on elastic substrata 
have been measured
very precisely \cite{alamo07,lombardi07}. The observed stresses range
up to $\sim$50Pa, giving rise to contractile pole forces, 
defined as the total force exerted in the front and back half of
the cell, of
$\sim$90pN. Typically, the contractile forces are concentrated in
spots of $\sim\mu$m size. These experiments also reveal a strong
correlation between force generation and morphological changes associated with the aforementioned three-stage cycle.  Thus, the cell
motion exhibits a mechanical cycle consisting of (i) a contraction
phase, initiated by pseudopod attachment, in which the stresses increase; (ii) a retraction phase, in
which the rear detaches and is brought forward. Consequently the cell
shrinks and the stresses relax; (iii) a protrusion
phase, in which the cell extends a pseudopodium in the direction of
motion. At this stage, the pseudopodium does not exert noticeable
forces on the substratum.  The length of such a cycle is in the order
of $\sim$1-2min for wild-type (WT) \textit{Dictyostelium} cells 
and $\sim$4min in cells lacking myosin II, 
a motor protein responsible for
cytoskeletal force generation \cite{alamo07}.  
The cell displacement of 15$\mu$m per cycle is
roughly constant.

The exact nature of the adhesive forces between \textit{Dictyostelium}
cells  and the substratum is not known.
Most likely, the observed forces are transmitted through discrete
contact foci on the ventral side of the cell. These foci are
associated with F-actin rich regions which appear in spatial and
temporal proximity to stress foci \cite{uchida04,iwadate08}. Actin
foci are spatially static but have a lifetime of $\sim$20sec.  Wild-type (WT)
cells have $\sim$5-10 foci.  On the other hand, based on experimental
results on cell detachment in shear flow, the number of microscopic adhesive
bridges between cell and substratum is estimated to be $\sim$10$^5$
\cite{simson98,decave02a}. Hence each adhesion focus is comprised of
many bridges.

It is reasonable to expect that to some extent, the cell speed should be controlled by the strength of attachment and the dynamics of detachment. Clearly, neither a non-adherent cell nor a cell
that is unable to detach can move.  However, between these extreme
cases, the cell speed seems to depend only weakly on
its adhesiveness~\cite{jay95}.  In support of this, weakly adherent
talin-null cells move with roughly the same speed as WT cells
\cite{alamo07}.  
Mutants lacking myosin II move more slowly than wild-type cells,
but cover the same distance per contraction cycle, i.e., the period of
the cycle is increased. These cells do exhibit a  much reduced
motility on strongly adhesive substrata~\cite{jay95}, as this combination places the cells in the extreme case of not having enough strength to contract against the adhesive forces. Finally, the over-expression of 
paxillin reduces the adhesion, but leaves the speed during folate chemotaxis relatively unchanged~\cite{duran09}.

The importance of attachment/detachment dynamics for cell motility has
been addressed in many theoretical studies
\cite{dimilla91,bottino98,gracheva04,larripa06}. These typically predict a strong
dependence of cell speed on cell-substratum adhesiveness. Indeed, the
prediction of an optimal adhesiveness is in excellent agreement with
experimental findings on mammalian cells \cite{palecek97}. But, as just discussed, the situation appears to be different in \textit{Dictyostelium}.

In these models, cell motion follows either from the protruding
activity at the front \cite{gracheva04,larripa06} or from asymmetric
detachment during cell contraction
\cite{dimilla91,bottino98,bottino02}.  In the latter models, cell
contraction is represented by internal forces acting on a
visco-elastic cell body and the
attachment/detachment dynamics are represented by an effective
friction term with the substratum \cite{bottino02,gracheva04,larripa06}. However, the
experimental observation of discrete binding sites suggest that a
representation by discrete, breakable springs as in
Refs. \cite{dimilla91,bottino98} is more appropriate. 

In this work we argue, that 
contraction takes place at a constant rate and
that the cell
speed is limited by the rate of detachment of the adhesive bridges.
That is, the rate-limiting step in cell motility in this case is the
peeling of the cell from the substratum.  Based on stress patterns observed in
Ref. \cite{alamo07} we assume that cell detachment takes place mainly
during the contraction phase and that protrusion forces contribute
only a small amount to cell detachment.  Therefore, our theoretical
model of cell motion emphasizes the role of cell detachment during the
contraction phase. Our model makes testable predictions about the cell speed under
various experimental situations. 
These include crawling on substrata
with varying adhesiveness and the variation of a number of cell-specific
parameters.

\section{Model}

\subsection{Components and Assumptions}

\begin{figure}
  \includegraphics[width=.5\textwidth]{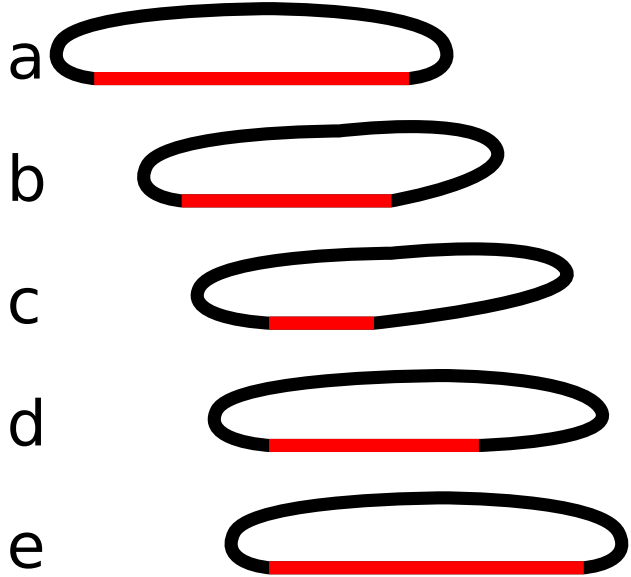}
  \caption{Schematic cross-section of a 
crawling \textit{Dictyostelium} cell illustrating the 
motility cycle. The part
    of the ventral surface that is in adhesive contact with the
    substratum is shown in red. (a) At the start of the contraction
    phase, the contact area is maximal. (b) During the
    contraction phase, the contact area shrinks while the
    cell is continuously transporting its body to the front. (c) At
    the end of the contraction cycle, the cell body is
    transported as far to the front as is allowed by the rear-most
    adhesions. Note, that it is assumed, that the protrusive force
    itself does not contribute significantly to the peeling of the
    rear. (d and e) During the relaxation phase, contraction stops and
a full ventral adhesion area is re-established beneath the
    pseudopodium.  }\label{fig:crawling}
\end{figure}

Our model focuses on the contraction phase of the motility cycle and 
does not explicitly treat the protrusive forward motion part of the cycle.
Instead, as is shown schematically in Fig. \ref{fig:crawling}, 
the cell is assumed to maintain protrusive activity 
throughout the cycle during which cell material 
is constantly transported to the front. 
This notion is corroborated by the observation that
over the entire cycle the cell speed shows only little variation (del
\'Alamo, private communication).  
Then, we can define the cell speed as the displacement of the back of the cell
at the end of the contraction phase divided by the cycle period.

We assume that during the contraction phase, 
with duration $\tau$, myosins contract the cell
body uniformly with a constant speed.  
This is motivated by (i) direct inspection of contracting 
\textit{Dictyostelium} cells \cite{alamo07} and (ii) the observation
that the \textit{in vitro} myosin velocity is load-independent \cite{riveline98}. 
The assumption of a constant contraction rate is an essential
difference to earlier work in which the 
cell is described as a one-dimensional network of
contractile elements, each of which is exerting the same force on the 
nodes of the network \cite{dimilla91}. 
Our choice is motivated by the fact that 
force balance implies that, when
attached elastically to a substratum, the interior of such networks is
largely stress-free. 
This is, however, in contrast to experimental observations which 
show that the stress field extends into the interior of the
cell-substratum area, indicating that cells do not operate
 a contractile network with prescribed forces.
We also assume that the cell contraction is not hindered by
viscous stress of the surrounding medium.
Indeed, as shown in Ref. \cite{alamo07},
the forces due to fluid drag on the moving cell are much smaller 
than the experimentally observed forces exerted on the substratum
($\sim$0.1pN vs. $\sim$90pN, \cite{alamo07}).
Thus, the cell is always in a
state of mechanical equilibrium and the motion of the cell is quasi-static.

Further, we assume that the cell is attached to the substratum via
adhesive bridges. These bridges form with a
fixed on-rate $k_+$ and dissociate with an off-rate $k_{-}$ which is
both force and position dependent.  The force dependence accounts for
the fact that the potential barrier between bound and unbound state is
lowered by an external force \cite{haenggi90,decave02a}. 
The position dependence incorporates a possible
preferred detachment at the rear vs. the front
\cite{dimilla91}.  
These asymmetric adhesion properties are known to play a 
major role in mammalian cells, where focal adhesion complexes are
coupled to intra-cellular pathways \cite{sabouri08}.
To our knowledge, and contrary to other 
systems \cite{lee97,huttenlocher97,kaverina02}, such a 
differential adhesion  has not been measured yet in  {\it
Dictyostelium}.

\subsection{Rigid Substratum Model}

\begin{figure}
  \includegraphics[width=\columnwidth]{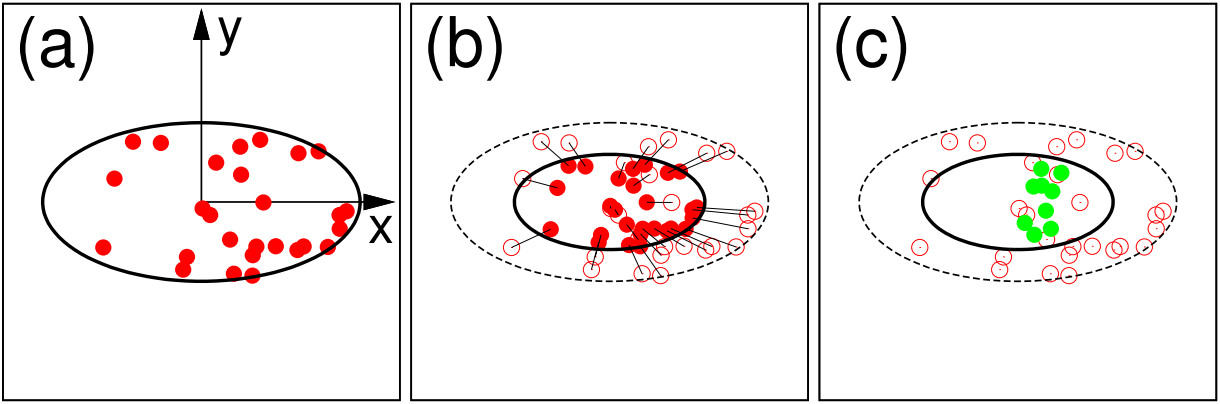}
  \caption{The cell-substratum contact area during different stages of
the contraction cycle. (a) The start of the contraction cycle with the
adhesion sites shown as red circles. The position of these sites is
measured in a coordinate system  with the center of the ellipsoid as 
the origin.
(b) During the contraction cycle the cell contracts uniformly at a 
constant speed. The initial position of the adhesion sites is shown as 
open circles while the current position is indicated by a solid
circle.
(c) The end of the contraction cycle,   with the remaining attached 
sites shown in green.   
    }\label{fig:model_sketch}
\end{figure}

In our simulations, the adhesion area is represented by an ellipse
with a fixed number ($N$) of randomly distributed sites that can adhere to the 
substratum.  Their position $\vec{x}_i(t)$ at time $t$ 
is measured with respect to the center of the ellipse.
The amount of contraction is parametrized by the contraction rate $\lambda$
which can take on values between 0 and 1 and which is defined as 
$\lambda=(R-R_{\tau})/R$ where $R, R_{\tau}$ are the semi-major axes
of the ellipse at the onset and end of contraction, respectively.
We divide the contraction cycle into 100 equal timesteps $dt$
and at each timestep the new position of node $i$ is given by 
$\vec{x}_i(t+dt)$=($\vec{x}_i(t)-\vec{x}_m(t))(1-\lambda dt / \tau)
+\vec{x}_m(t)$.
Here, $\vec{x}_m(t)$ is the location of the cell's center which 
is allowed to shift in order to ensure a vanishing net force on the
cell (see below). 
The position dependence of the off-rate is chosen to depend on the 
component $x$ along the direction of motion as follows:
\begin{equation}
  k_-(x)=k_{-,b}-[k_{-,b}-k_{-,f}]\frac{x-x_b}{x_f-x_b}\ ,
\end{equation}
where $x_{f/b}$ represent the front/back of the cell at the start of
the contraction cycle and where $k_{-,f/b}$ 
are independent parameters of our model.
The probability that a particular site adheres is 
given by the equilibrium value $\frac{k_+}{k_++k_-(x)}$.

The attachments between cell and substratum are modeled by elastic
springs with spring constant $k_s$. 
In the case of a very rigid substratum we can ignore the deformations
in the substratum. 
Then, the force on a
single bond is given by $\vec{F}_i(t)=k_s(\vec{x}_i(t)-\vec{x}_i^0)$, 
where $\vec{x}_i^0$ is the initial position of the bond.
In principle, our prescribed displacement of the nodes can lead to a non-zero
net force on the cell. 
To ensure a vanishing net force after each 
iteration we use the fact that the motion is quasi-static and 
allow the ellipse to shift and rotate. 
Specifically, we minimize
the total energy of the springs at time $t+dt$,
\begin{equation}
  E_s =
  \frac{k_s}{2} \sum_i\left[R_\varphi(\vec{x}_i(t)-\vec{x}_m(t)) +
  \vec{x}_m(t+dt) - \vec{x}_i^0 \right]^2\
  \label{eq:total energy}
\end{equation}
where $R_\varphi$ is the matrix describing a rotation by
$\varphi$. 
An implementation of this minimization procedure revealed
that the shift of the cell's center is small ($<$ 5\%) compared 
to the translation of the cell for most of our model parameters 
and only became significant for small ratios of $k_{-,f}/k_{-,b}$.
To compute the resulting traction stress, $\sigma$, we
tile the substratum into 0.05 $R$$\times$0.05 $R$ squares and 
compute the total force per area for each tile.

The force dependence of the off-rate is
approximated by an exponential factor \cite{bell78},
\begin{equation}
  k_-(\vec{x}_i(t))=k^{(0)}_-(\vec{x}_i^0)
  \exp\left(\alpha\frac{|\vec{x}_i(t)-\vec{x}_i^0|}{R}\right),
  \label{eq:off_rates}
\end{equation}
where we have defined the dimensionless parameter 
$\alpha\equiv R k_s\Delta/(k_bT)$.
The molecular length scale $\Delta$
characterizes the width of the potential well which prevents the
adhesive bridge from breaking and is of the 
order of 1 nm \cite{bell78}.

Attachment of bridges to the substratum is assumed to occur with a
force-independent rate constant $k_+$. Binding rates decrease
exponentially with the distance between membrane and substratum
\cite{decave02b}. Therefore we assume that attachment occurs only
inside the contracted ellipse. We assume that 
$k_+$ is uniform across the contact area. 
The density of bridges on the membrane
is assumed to be constant, such that the total number of available
bridges that can attach
at time $t$ is proportional to the area of the contact area 
$ \sim N(1-\lambda t/ \tau)^2$.

The uniform contraction builds up stress and, consequently, 
a number of foci will detach
during the contraction phase.
To calculate the speed of the cell we first compute the 
smallest value of the $x$-component for all attached foci, 
$x_{min}(0)$, at the start of the 
contraction cycle. This corresponds to the left-most attachment
point in Fig. \ref{fig:crawling}a. 
At the end of one contraction cycle, we determine the focus 
with the smallest value of the $x$-component, $x_{min}(\tau)$
(left-most point in  Fig. \ref{fig:crawling}c). 
Then, the speed of the cell is given by 
$(x_{min}(\tau)-x_{min}(0))/\tau$.    
For each parameter set, we performed 1000 independent contraction 
cycles and parameter values were chosen such that there is
at least 1 remaining  attachment point.

\subsection{Elastic Substratum Model}

Traction force experiments that measure the position of fluorescent 
beads require the use of deformable
substrata. The observed deformations are typically $\sim0.2\mu$m
\cite{alamo07}, comparable to the typical length of
adhesion molecules \cite{marshall06}. Under these conditions, the
adhesive bridges cannot be treated as non-interacting springs. Rather,
the elongation of a bridge under a prescribed cell contraction is
influenced by the amount of substratum deformation caused by neighboring
springs.  

To capture this effect, we simulated  a deformable substratum with Young's
modulus $E$ as a two-dimensional triangular network of springs with
spring constant $k_{sub}$ and rest length $L$.  
In these simulations, the initial conditions, the on and off rates 
of the cell nodes and the
contraction procedure are the same as
described above. 
Now, however, we need to compute the new positions of 
the triangular mesh vertices after each timestep. 
For this, we compute the total energy, given by
\begin{equation}
  E(t) = 
  \frac{k_{sub}}{2} \sum_{i,j} ( | \vec{y}_i(t)-\vec{y}_j(t)| - L )^2 
  +
  \frac{k_s}{2} \sum_{a=1}^{N_a} ( \vec{x}_a(t) - \vec{y}_{i_a}(t) )^2 \ .
  \label{eq:energy elastic}
\end{equation}
Here, $\vec{y}_i(t)$ is the position  of the i-th 
triangular mesh vertex at time $t$.
The first sum in Eq. \ref{eq:energy elastic} extends
over all pairs of neighbors in the triangular grid and the second
sum runs over the substratum nodes that
are coupled to $N_a$ adhesive springs.
For simplicity, we have chosen boundary conditions in which 
the position of the substratum boundaries are fixed. 
Minimization of Eq. \ref{eq:energy elastic} directly yields the
new positions of the vertices and, thus, the
deformation pattern of the substratum. 
The force exerted on the
attachment point $\vec{y}_{i}(t)$  by the cell node $\vec{x}_j(t)$ 
can be calculated as $ \vec{F}^j(\vec{y}_{i},t)=k_s
(\vec{x}_{j}(t) - \vec{y}_{i}(t) )$.
The total force $\vec{F}(\vec{y}_{i},t)$
on each attachment point is then the sum 
over all nodes $j$ connected to this point. 
These point forces are related to the local applied stress via
\begin{equation}
  \sigma_{\nu z}(\vec{y}_{i},t)
  =
  \frac{2}{\sqrt{3}}\frac{F_\nu(\vec{y}_{i},t)}{L^2}\ ,
  \quad \nu=x,y\ .
\end{equation}

Note that our choice for the boundary condition 
will lead to non-zero net forces on the cell. 
We found, however, that for a substratum of sufficient size (4Rx4R)
the net force is less then 5\% of the pole force.
Of course, by repositioning the cell after each time step we 
could ensure a vanishing net force even in the case of fixed boundaries.
Furthermore, choosing periodic boundary conditions for the substratum
will also guarantee a vanishing net force on the cell.
We found that the resulting force pattern differs only 
slightly from the force pattern
generated using fixed boundaries, demonstrating that the results are 
insensitive to the precise details of the numerical algorithm.

\subsection{Parameter Estimates}

Throughout the paper we will use a default set of parameters that
were obtained, where possible, from experimental data.
The shape of the cell is characterized by a long semi-axis, taken to 
be $R=$10$\mu$m, and an aspect ratio 1:4.  Based on movies shown as
supplemental material to Ref. \cite{alamo07} and direct measurements
of the adhesion area in Refs. \cite{schindl95,weber95}, we assume that
the (WT) cell contracts by 50\% of its length, 
corresponding to $\lambda=0.5$ in our simulations, during a
contraction period of $\tau$=1 min. 

For the number of adhesive bridges we followed Refs.
\cite{decave02b} and chose $N=200$. 
Note, however, that our results do not depend on $N$ as long 
as we rescale the other model parameters appropriately. 
Specifically, if $N \rightarrow \mu N$ we need to rescale $k_s$ and 
$\Delta$ as follows: $k_s \rightarrow k_s/\mu$ and $\Delta \rightarrow
\mu \Delta$.
The off-rates are estimated in models of shear flow
induced detachment \cite{decave02a,decave02b}  and at the back we take 
$k_{-,b}=$ 1 $\times 10^{-2}$/sec. 
As discussed before, there is no clear
data on the possible maturation of adhesion sites
in {\it Dictyostelium} and we have arbitrarily chosen 
the off-rates at the front to be  equal to 0.5$k_{-,b}$. 
The force dependence of the off-rate in
Eq. \ref{eq:off_rates} is determined by the dimensionless parameter 
$\alpha$ which we have chosen to be 125.
This parameter is a combination of the rupture width $\Delta$ of
the molecular bond and the adhesive spring constant $k_s$. 
We have chosen the latter to be $k_s= 1 \times 10^{-4}$N/m,  which is 
in the range of  experimental values  
\cite{marshall06}, and  $\Delta\sim0.5$nm \cite{bell78}.
Finally, the spring constant of the deformable substratum was
estimated using the experiments results in Ref.  \cite{alamo07}.
There, the pole force was found to be $F_p\sim$ 200 pN while the 
deformation was $u\sim 0.2 \mu$m, leading to 
$k_{sub}=F_p/u=1 \times 10^{-3}$N/m.

\section{Numerical Analysis and Results}

\subsection{Rigid substratum}

\begin{figure}
  \includegraphics[width=.5\textwidth]{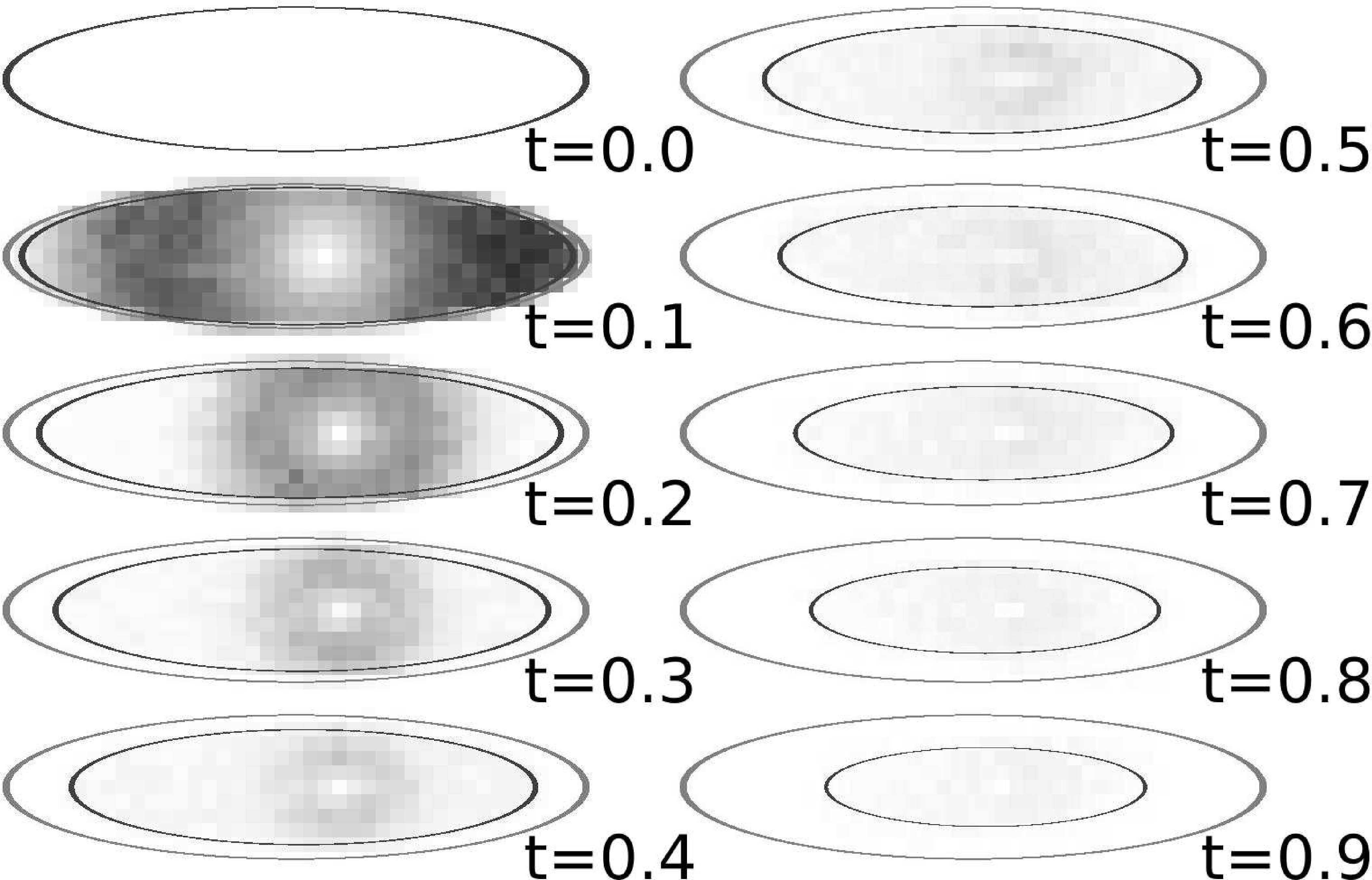}%
  \caption{ Average traction stress patterns over 1000
    simulation runs with time expressed in units of the contraction
cycle. For the purpose of averaging the distribution
    maps were tiled into 0.05 $R$$\times$0.05 $R$
    squares. The stress is shown in a  gray-scale with 
    black corresponding to a traction stress of
    $\sigma \approx$4.4$k_sR^{-1}$. At the beginning of the contraction cycle
    ($t$=0) no force is exerted. The outer ellipse indicates the
    original position of the cell and the inner ellipse indicates the
    current adhesion area. 
    }\label{fig:frames}
\end{figure}

With the above choice of parameters, we performed 
1000 contraction cycle simulations.
At time intervals $dt=0.01\tau$ the distribution of
displacements $\vec{u}_i=\vec{x}_i-\vec{x}_i^0,\,i=1...N$ was stored.
The displacements $u_x$ and $u_y$ are directly related to the traction
forces exerted on the substratum via $F_i=k_su_i,\ i=x,y$.
Fig. \ref{fig:frames} shows the time evolution of the stress
averaged over 1000 individual runs. Here, forces were
summed up in bins of size 0.05 $R$$\times$0.05 $R$.
Note, that the ellipses in our simulations correspond to the adhesion
area which does not necessarily correspond to the 
experimentally determined cell outline \cite{alamo07,lombardi07}. 

\begin{figure*}
  \includegraphics[width=.5\textwidth]{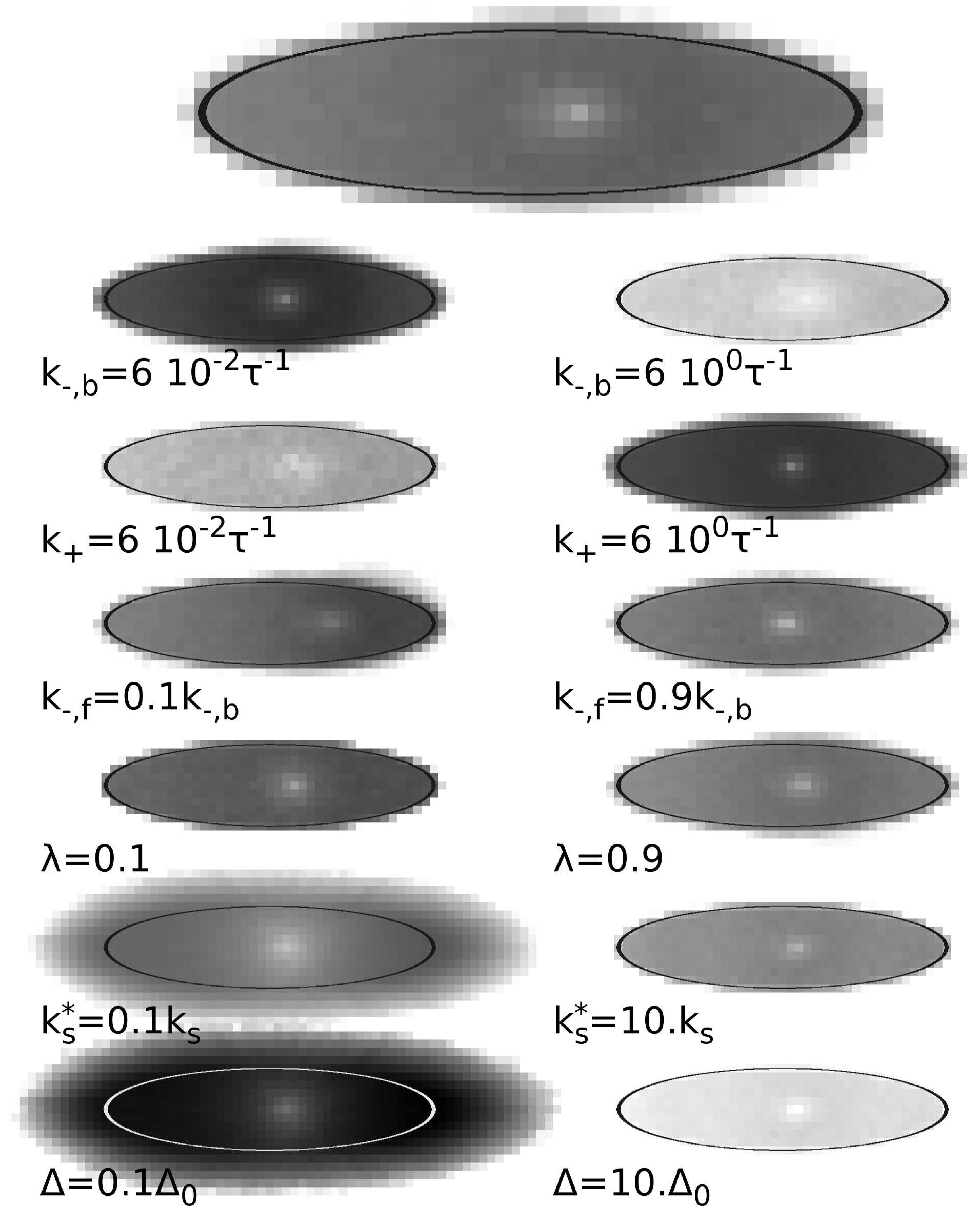}
  \caption{ The traction stress $\sigma$, averaged over an entire 
contraction cycle, for different sets of model
parameters.  
The stress is plotted using a logarithmic gray scale
with black corresponding to $|\sigma|\approx6.5k_sR^{-1}$ and white
    corresponding to values $|\sigma|<6.5\cdot10^{-3}k_sR^{-1}$. 
    The time averaging was achieved by rescaling and overlaying
the contracted ellipses. 
The upper pattern corresponds to
the default set of parameters:
    $k_{-,b}$=6$\cdot$10$^{-1}\tau^{-1}$, $k_{-,f}$=0.5$k_{-,b}$,
    $k_+$=6$\cdot$10$^{-1}\tau^{-1}$, $\alpha$=125, $\lambda=0.5$, and $N=200$.  For
    this set of parameters, the maximal stress is
    $\approx$0.6$k_sR^{-1}$. In each row one model parameter is
    varied while keeping the remaining parameters fixed. 
  }\label{fig:comparison}
\end{figure*}

Fig. \ref{fig:comparison} shows the force distribution averaged over
time for different choices of model parameters.  Averaging was done by
scaling individual time frames such that the
\textit{contracted} ellipses fall on top of each other.  
The top pattern corresponds to the average stress pattern 
for the default parameters.
In each row of images, we have varied one of these parameters 
and have plotted the stress pattern using a gray-scale with 
black corresponding to large stresses.
Surprisingly,
variations in the amount of contraction $\lambda$ and the relative
adhesiveness $k_{-,f}/k_{-,b}$ have only little influence on the
stress pattern. Rather, it depends more strongly on the
\textit{basal} detachment/attachment dynamics via $k_{-,b}$ and $k_+$, 
along with the molecular length scale $\Delta$ and
the spring constant $k_s$.  

\begin{figure}
  \includegraphics[width=.45\textwidth]{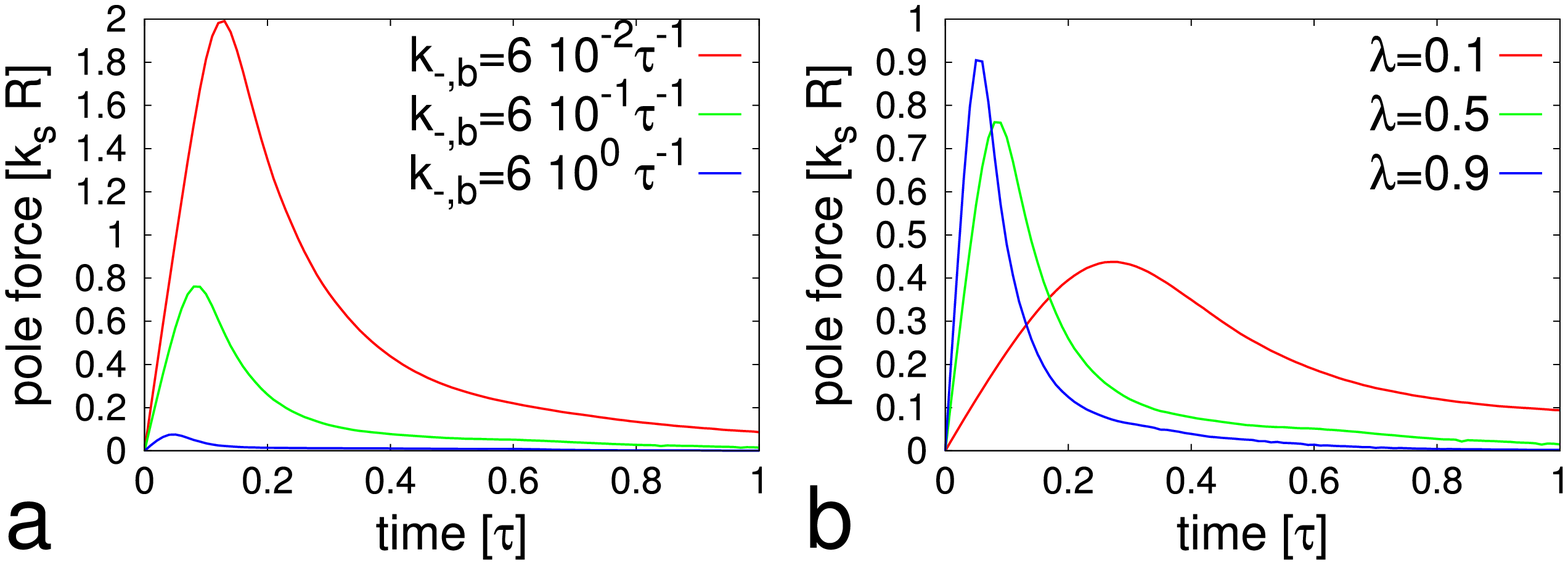}%
  \caption{Average pole-forces as function of time during one
    contraction cycle. The default parameter set 
is used and the parameter value indicated in the legend is varied. 
    }\label{fig:energy}
\end{figure}

In Fig. \ref{fig:energy} we plot the dependence of the pole forces
on the model parameters as a function of time during 
a contraction cycle. The pole force at the
back, $F_b$, is defined as the total force exerted in the direction of
motion, i.e.
\begin{equation}
  \vec{F}_b\equiv k_s\sum_{u_x>0}\vec{u}\ .
\end{equation}
Similarly, the pole force at the front, $\vec{F}_f$, comprises all
forces which point into the negative $x$-direction. 
Our definition of the pole forces differs from the one in  
Ref. \cite{alamo07}, where pole forces are  defined as the
overall forces transmitted at the attachment regions in the front and
back halves of the cell.
In each graph we used the default parameter set and varied one 
value as indicated in the legend.

\begin{figure}
    \includegraphics[width=.45\textwidth]{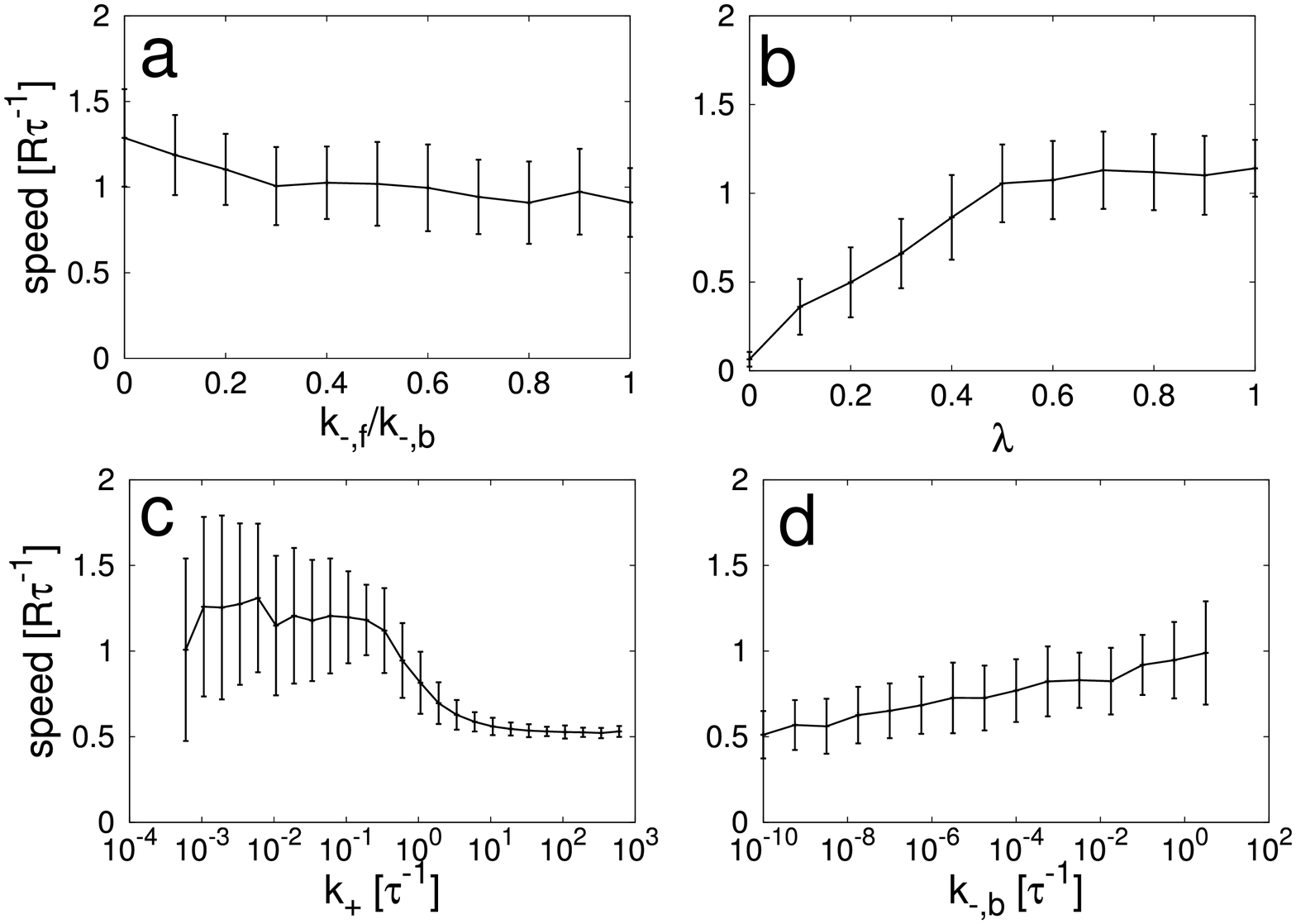}%
  \caption{ The dependence of the cell speed on one out of the six
    model parameters is shown. The remaining parameters are fixed at
their default values.
}\label{fig:scan}
\end{figure}

\begin{figure}
  \includegraphics[width=.45\textwidth]{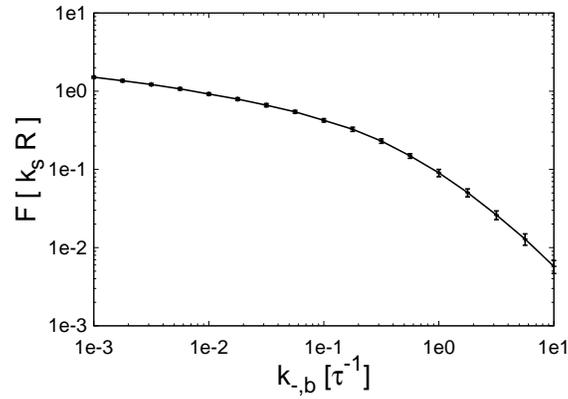}%
  \caption{ The  average pole force exerted during one contraction
    cycle as a function of $k_{-,b}$. 
  }\label{fig:scan_koff_frc}
\end{figure}

In Fig. \ref{fig:scan} we compare the dependence of the cell speed on 
four model parameters. In our model, this speed is determined by the
amount of retraction at the rear of the cell per contraction cycle:
during the protrusion phase the ellipse representing the 
protruded cell outline is moved such that the rear coincides with 
the last remaining attached focus.
The actual forward motion is
accomplished throughout the contraction and protrusion phase (see Fig.
\ref{fig:crawling}).
Note that even for a symmetric
detachment the cell can move forward. 
Again, we varied one parameter value with the remaining parameters 
fixed at the default values.
Finally, in Fig. \ref{fig:scan_koff_frc} we plot the  
average pole force during a single motility cycle as a function of 
the off-rate $k_{-,b}$. As expected, 
the  pole force decreases as the 
off-rate increases. 

\subsection{Elastic Substratum}

\begin{figure}
  \includegraphics[width=.45\textwidth]{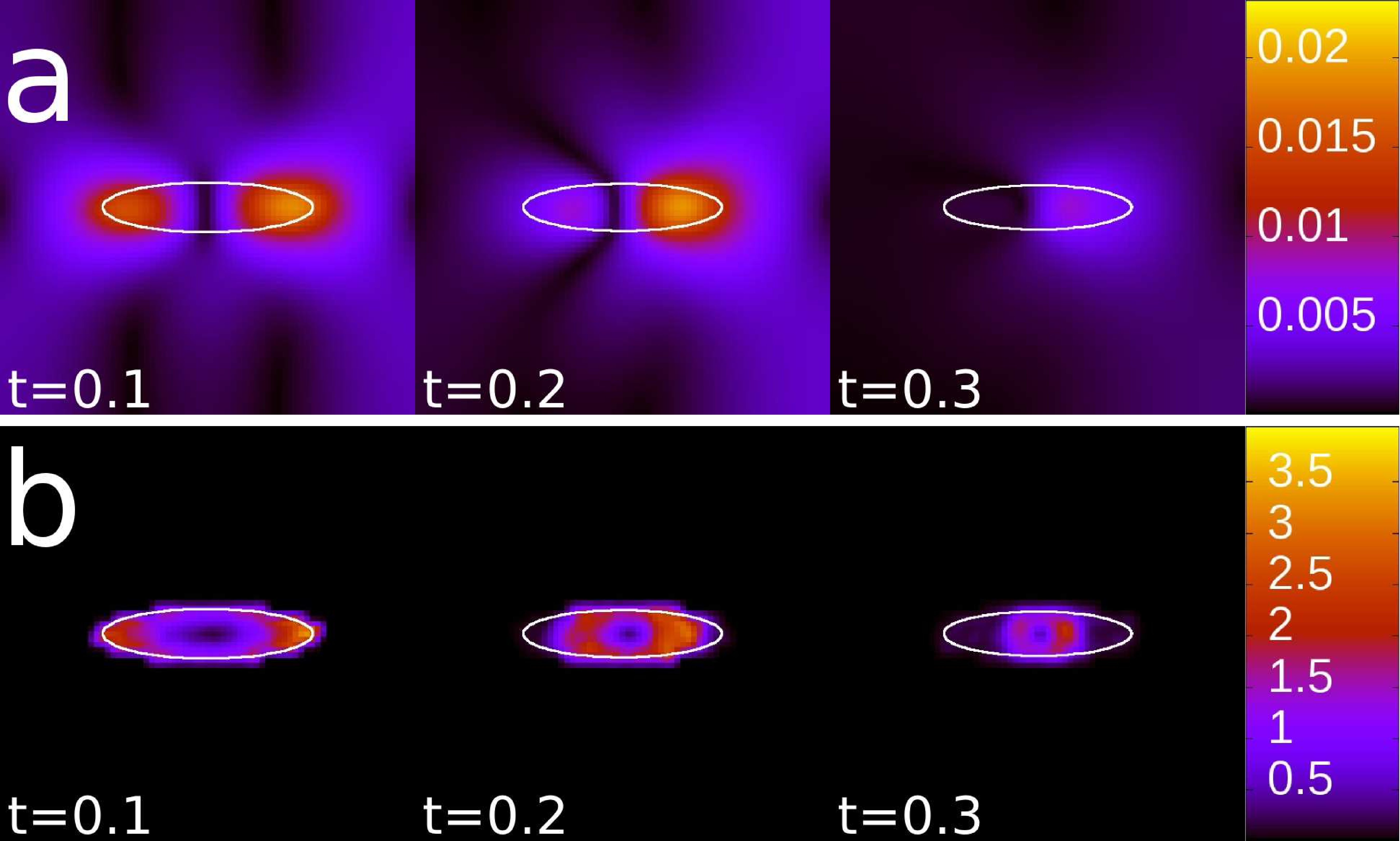}%
  \caption{Time evolution of the displacement pattern (a) and 
stress pattern (b) of a deformable
    substratum. Displacements are given in units of $R$ using 
the displayed color scale while stresses are given in $k_s/R$, as
shown in the color scale. In this
    simulation, the default parameter values were used 
together with an effective spring constant $k_{sub}=10 k_s=
1 \times 10^{-3} N/m$.
    }\label{fig:displacements}
\end{figure}

Fig. \ref{fig:displacements}a shows a time series of the displacement
pattern for an elastic substratum. The displacement is shown using
the indicated color scale and 
the computed maximal displacements $\sim0.02R\sim0.2\mu$m are in good
agreement with experimental results. Due to rapid detachment,
substratum deformations vanish shortly after onset of
contraction.
Fig. \ref{fig:displacements}b shows the corresponding stress distribution
$(\sigma_{xz}^2+\sigma_{yz}^2)^{1/2}$ 
which shows two
distinct peaks at the front and the back corresponding to regions of
maximal displacement in Fig. \ref{fig:displacements}a.  The computed
maximal stresses ($\sim4k_sR^{-1}\sim40$Pa)  
are similar to the ones observed
in experiments ($\sim50$Pa \cite{alamo07}).

\section{Discussion}

In this paper we have presented a mathematical model for cell motility  
motivated by experimental observations of the motion of 
{\it Dictyostelium} cells. The emphasis of our model is on 
the interaction between the cell membrane and the substratum
on which the cell is crawling while the actual cell deformation and 
translation are not explicitly taken into account. 
There are several distinct differences between our approach 
and previous modeling studies.
The studies carried out by Lauffenburger and co-workers \cite{dimilla91}, for 
example, considered a one-dimensional cells with only a handful 
of attachment points. These points were connected through springs
that are exhibiting a prescribed force.
In our model, on the other hand,
the foci are moving with a constant contraction rate. 
This choice was motivated in part by the observed stress and force
patterns in traction force experiments. These experiments 
demonstrate that the forces are maximal within the 
contact area. 
In a model where the inter-foci springs  exert a fixed 
force, the force field within the interior of the 
contact area will be very small and concentrated 
at its boundary. 
Furthermore, experiments on TalinA- cells \cite{alamo07} demonstrate
that cells with a vastly reduced adhesion move with roughly 
the same velocity as wild-type cells. 
A prescribed force model would predict a strong dependence of the 
cell's speed on the adhesion  strength.

Another major difference is the presence of dashpots,
representing the viscous nature of the cell's 
cytoplasm, in the earlier models. 
These dashpots play an important role if one prescribes the force
exerted by contractile elements.
Here, however, we prescribe the contraction velocity which
alleviates the need for an explicit modeling of the viscous cytoplasm.
The estimation of the contraction speed we used in our simulations,
$\sim 10 \mu$m/min, is based on direct experimental observations.  
However, typical \textit{in vitro} myosin
velocities measured in motility assays are $\sim 10-20$ times higher than
the experimentally observed cell speeds \cite{riveline98,murphy01}. 
The \textit{in vivo} velocity is not known but 
will likely be of the order of the contraction speed. 
The mechanism responsible for this significant slow down 
is, to our knowledge, unclear. 
One possibility is that \textit{in vivo} the disordered structure of 
the actin-myosin cortex hinders a rapid contraction. 
Also, the viscosity of the cytoplasm may play an
important role in limiting the myosin contraction speed.  
A final difference is that 
our two-dimensional model explicitly takes into account the 
displacement of the rear. 

Another class of models describe the cell as a gel,
with visco-elastic properties \cite{bottino02,gracheva04,larripa06}.
Contrary to our model, these studies prescribe the protrusion 
of the cell and do not focus on the contraction mechanism. 
In these models, the adhesion has a front-to-back 
gradient and is represented by an effective friction force.
Thus, they are unable to address the role of contraction on
the detachment of crawling cells. 

Most of our results are obtained assuming that the 
substratum is rigid, corresponding to a typical experimental
set-up where cells are crawling on glass surfaces. 
In this case, the displacement of the adhesion proteins
is much larger than the displacement of the 
attachment point at the substratum.  Thus, the force field 
exerted on the substratum is simply determined by 
the forces on the adhesion proteins. As expected, this
average  traction force  varies during the contraction
cycle and reaches its  maximum shortly after the start
of the cycle (Fig. \ref{fig:frames}). 
The pattern observed in Figs. \ref{fig:frames} and 
\ref{fig:comparison} can be explained by realizing that 
in our model stress is generated by a prescribed isotropic
contraction.  This leads to radial increase of stress at the adhesions
which, in the absence of binding/unbinding dynamics, is given by the
geometry of the contraction only. 
Thus, in our model the binding sites at the center of the
adhesion zone are always almost stress free, resulting in the 
observed pattern. 

The default set of parameters of our model were based, 
where possible, on experimental values. 
To examine the effect of these parameters on the 
force patterns, we have systematically changed one 
while keeping the remaining parameters fixed (Fig.
\ref{fig:comparison}).
The stress  pattern depends strongly 
on the molecular length scale $\Delta$
with the stress increasing for 
smaller values of $\Delta$.
This parameter determines the
off-rate  of the bridges (Eq. \ref{eq:off_rates}) and 
for small values of $\Delta$, this rate 
becomes small. Correspondingly, 
the force per focus becomes large, leading to large stresses 
shown in Fig.  \ref{fig:comparison}. 
Note that the parameter $k_s$ also controls the off-rate. 
A change in $k_s$, however, does not change the force pattern
as dramatically as a change in $\Delta$ 
since this parameter  determines the force per bridge
as well.

The relative adhesiveness $k_{-,f}/k_{-,b}$  measures the asymmetry 
in the adhesion strength between the front and the back of the cell. 
Such an asymmetry is essential to the motility of 
mammalian cell but  its role in {\it Dictyostelium}  movement is 
unclear.
Variations in the amount of the relative
adhesiveness $k_{-,f}/k_{-,b}$ have only little influence on the
magnitude of the observed stress pattern. 
The pattern, however, becomes more asymmetric as 
$k_{-,f}/k_{-,b}$ decreases. 
Clearly, a larger off-rate at the front than at the 
back will lead to a higher concentration of attached bridges  at the 
front and thus a larger stress  in 
the front half of the cell.

The parameter $\lambda$ describes the amount of contraction. 
In the absence of detachment, a larger contraction would lead to
an increase in the elongation of the bridges and   
a larger force per area. 
However, the increased force on the foci will lead to an 
increase in the detachment and, 
as can be seen from Fig. \ref{fig:comparison}, these 
two effects compensate and lead to a slightly  smaller
time averaged stress for larger contractions.

The on-rate $k_+$ describes the re-attachments of foci and 
increasing the value of $k_+$ will result in an increase
in the number of attached foci during the contraction 
cycle. Thus, the force per area  increases for increasing 
values of $k_+$, as is evident from Fig. \ref{fig:comparison}.
The off-rate  $k_{-,b}$, on the other hand, determines the 
detachment dynamics of the foci. 
A higher value of $k_{-,b}$ leads to a smaller number of 
attached foci and thus a smaller force per area. 

The pole forces, defined as the sum of all the forces 
parallel  or anti-parallel to
the direction of the motion, increase rapidly  and 
linearly at the start of the 
contraction cycle (see Fig. \ref{fig:energy}). 
This linear behavior can be understood by realizing that 
during the initial contraction period, the force dependence of
the off-rates is insignificant and the number of
bridges stays roughly constant. Since the force on each adhesion is
proportional to the contraction ratio, the pole force increases
linearly.
Once force induced detachment becomes significant
the bridges begin to break and the pole force starts to decrease.
The maximum pole force, and the time at which 
this maximum is reached, depend on the 
model parameters (Fig. \ref{fig:energy}). 
In particular, 
the maximum value increases for smaller values of $k_{-,b}$
(Fig. \ref{fig:energy}a).
After all, small values of the off rate lead to larger
displacements and, thus, larger forces.
Furthermore, the pole force increases for
larger values $\lambda$ (Fig. \ref{fig:energy}b) which 
can be understood by realizing that  
small contractions lead to small displacements and thus 
smaller pole forces. 

Using our model, we are able to vary systemically 
each parameter and determine the dependence of the speed on this 
parameter. 
The results (Fig. \ref{fig:scan}) can be viewed as experimental 
predictions even though we realize it might be difficult to 
vary some of these parameters in experiments.
In particular, it is not always obvious which adhesion parameter 
is probed in a certain experiment and how the parameters are
changed in a certain mutation. 
For example, the reduced adhesiveness of TalinA- mutant 
may result from an increased off-rate or 
from a smaller total number of adhesive sites. 
Surprisingly, we find that the speed is only weakly dependent on the
relative adhesiveness $k_{-,f}/k_{-,b}$.
This is in contrast to previous models  where the speed 
depends critically on this ratio. 
Our model assumes that the  protrusion is decoupled from the 
contraction cycle (Fig. \ref{fig:crawling}). 
Thus, our speed is mainly determined by the peeling velocity of 
the back and can be significant even  for uniform off rates.
Note that for small relative adhesiveness it becomes important 
to ensure  a vanishing net force through a re-orientation of the 
cell outline. Without this re-orientation the cell's speed would be 
purely given by the off-rate at the back and would be constant for 
all values of the relative adhesiveness.  

As expected, we find that the cell speed increases for 
increasing values of the contraction rate $\lambda$ 
(Fig. \ref{fig:scan}b).
After all, in the limit of vanishing contraction rate the speed 
approaches zero while for maximal contraction rate the speed 
reaches a maximum.
Furthermore, we find that high on-rates decrease the 
speed (Fig. \ref{fig:scan}c). For high values of 
$k_+$, adhesive bridges are deposited at rates that are 
higher than the detachment rates, limiting the cell's speed.

Contrary to previous studies, we find that the 
speed does not depend strongly on the off-rate $k_{-,b}$ 
(Fig. \ref{fig:scan}d). 
Of course, the speed will approach 0 for 
very small values of this off-rate where the foci will remain attached
to the substratum. In this limit, we expect that our constant 
contraction speed assumption is no longer valid and that the 
forces on the myosin motors are large enough to lead to 
stalling.
For large values of the off-rate, all foci will detach and 
we have only considered the range of values for which 
at least one focus remains attached. 
In fact, in this limit the weakly adherent cells can exert only small
forces on the substratum, see Fig. \ref{fig:scan_koff_frc}. 
Hence, for sufficiently
large $k_{-,b}$, the traction force that balances the viscous drag of
the protruding cell ($\sim0.1$pN \cite{alamo07}) exceeds
the detachment force. For approximately symmetric cells, force balance
then implies that a forward protrusion is accompanied by a backward
motion of the same order. Hence, there is no net motion for
sufficiently large $k_{-,b}$. For the parameter range studied, the
traction force is always sufficient to support protrusive forward
motion, see Fig. \ref{fig:scan_koff_frc}.

Our finding that  the cell speed is roughly constant 
for a large range of values of adhesive forces is 
in agreement with recent experiments in which the stress
patterns of crawling {\it Dictyostelium} cells were examined. 
These experiments show that the cell motion can be described 
by a contraction-relaxation-protrusion cycle. 
Thus, the cell's speed is determined by the ratio of the displacement
per cycle and the period of this cycle.
TalinA- cells exhibit a drastically reduced cell-substratum
adhesion but were found 
to have the same 
cell speed as wild-type cells, with an identical period and, thus, identical
displacement. 
Of course, two data points cannot rule out a significant 
dependence of the cell speed on the adhesion strength and a 
definitive test of our model would be to examine the cell speed for 
different mutants. 
One candidate would be cells in which the expression level of PaxB,
the {\it Dictyostelium} orthologue of paxillin, is 
altered. Both PaxB- cells 
\cite{bukharova05} 
and cells  in which 
PaxB is overexpressed 
\cite{duran09} 
exhibit 
a decrease in cell-substratum adhesion.
The cell speed in cAMP gradients is reduced in 
PaxB overexpressed cells and is increased in PaxB- cells.
Interestingly, the cell speed in folate gradients is largely
independent of the expression level of PaxB 
(\cite{duran09} and D.  Brazill, personal communication).
This might indicate a PaxB role in the  periodicity of the 
motion cycle, which would effect the cell's speed. 
A more detailed analysis of these mutants that can 
measure force patterns and motility cycles would be 
interesting. 

A quantitative comparison with the experimentally obtained stress 
patterns is only possible if we take into account a deformable
substratum. After all, these experiments measure the displacement 
of fluorescent beads embedded in the substratum and 
require significant movement of these beads. 
Thus, our model assumption that the displacement of the substratum
is negligible compared to the stretching of the adhesive
bonds is no longer valid. 
To compare to experiments, we have extended our model and have
explicitly simulated a triangular spring network, representing 
the substratum. This extension renders the simulations 
computationally more demanding and we have only  performed
a limited set of simulations (Fig. \ref{fig:displacements}). 
Using experimental values characterizing the substratum, we 
found that our results show a quantitative and 
qualitative agreement with 
the experimentally observed stress and strain patterns.
For our experimentally based parameter values we obtained
a maximum displacement 
that was comparable to the one observed in 
experiments ($\sim  0.2 \mu$m). Furthermore, the computed  peak stress
is similar to the experimental peak stress: $~\sim$40 Pa vs. 
$~\sim$50 Pa. 

In summary, we have presented a simple model for the motion of 
{\it Dictyostelium} cells. We have shown that this model can 
produce a number of experimentally verifiable predictions and 
can be extended to include deformable substrata. 
Our strongest prediction, that the cell speed is largely independent 
of the value of the adhesive forces, should be testable 
using force cytometry experiments.
Our model focused on the cell-substratum interaction and 
ignored the protrusion phase of the motility cycle. 
Extensions that include intra-cellular signaling pathways that drive 
cell deformations are currently  under investigation.

This work was supported by the National Institutes of Health Grant P01
GM078586.  LMS was partly supported by NSF grant DMS 0553487, and
would like to thank the Center for Theoretical Biological Phyiscs for
hospitality.  MB gratefully acknowledges support from a German
Academic Exchange Service (DAAD) fellowship.  We thank Juan C. del
\'Alamo and William F. Loomis for useful discussions.


\begin{thebibliography}{39}
\providecommand{\url}[1]{\texttt{#1}}
\providecommand{\urlprefix}{ }

\bibitem[Franz et~al.(2002)Franz, Jones, and Ridley]{franz02}
Franz, C.~M., G.~E. Jones, and A.~J. Ridley, 2002.
\newblock Cell Migration in Development and Disease.
\newblock \emph{Dev. Cell} 2:153 -- 158.

\bibitem[Baggiolini(1998)]{baggiolini98}
Baggiolini, M., 1998.
\newblock Chemokines and leukocyte traffic.
\newblock \emph{Nature} 392:565--568.

\bibitem[Wang et~al.(2005)Wang, Goswami, Sahai, Wyckoff, Segall, and
  Condeelis]{wang05}
Wang, W., S.~Goswami, E.~Sahai, J.~B. Wyckoff, J.~E. Segall, and J.~S.
  Condeelis, 2005.
\newblock Tumor cells caught in the act of invading: their strategy for
  enhanced cell motility.
\newblock \emph{Trends Cell Biol.} 15:138--145.

\bibitem[Condeelis et~al.(2004)Condeelis, Song, Backer, Wyckoff, and
  Segall]{condeelis04}
Condeelis, J., X.~Song, J.~M. Backer, J.~Wyckoff, and J.~Segall, 2004.
\newblock Cell Motility, chapter Chemotaxis of Cancer Cells during Invasion and
  Metastasis, 175--188.

\bibitem[Kedrin et~al.(2007)Kedrin, van Rheenen, Hernandez, Condeelis, and
  Segall]{kedrin07}
Kedrin, D., J.~van Rheenen, L.~Hernandez, J.~Condeelis, and J.~Segall, 2007.
\newblock Cell Motility and Cytoskeletal Regulation in Invasion and Metastasis.
\newblock \emph{J. Mammary Gland Biol. Neoplasia} 12:143--152.

\bibitem[Rafelski and Theriot(2004)]{rafelski04}
Rafelski, S.~M., and J.~A. Theriot, 2004.
\newblock Crawling Toward a Unified Model of Cell Motility: Spatial and
  Temporal Regulation of Actin Dynamics.
\newblock \emph{Ann. Rev. Biochem.} 73:209--239.

\bibitem[Mogilner(2009)]{mogilner09}
Mogilner, A., 2009.
\newblock Mathematics of cell motility: have we got its number?
\newblock \emph{J. Math. Biol.} 58:105--134.

\bibitem[Lauffenburger and Horwitz(1996)]{lauffenburger96}
Lauffenburger, D.~A., and A.~F. Horwitz, 1996.
\newblock Cell Migration: A Physically Integrated Molecular Process.
\newblock \emph{Cell} 84:359 -- 369.

\bibitem[Lee and Jacobson(1997)]{lee97}
Lee, J., and K.~Jacobson, 1997.
\newblock {The composition and dynamics of cell-substratum adhesions in
  locomoting fish keratocytes}.
\newblock \emph{J. Cell. Sci.} 110:2833--2844.

\bibitem[Laukaitis et~al.(2001)Laukaitis, Webb, Donais, and
  Horwitz]{laukaitis01}
Laukaitis, C.~M., D.~J. Webb, K.~Donais, and A.~F. Horwitz, 2001.
\newblock {Differential Dynamics of {alpha}5 Integrin, Paxillin, and
  {alpha}-Actinin during Formation and Disassembly of Adhesions in Migrating
  Cells}.
\newblock \emph{J. Cell. Biochem.} 153:1427--1440.

\bibitem[Kaverina et~al.(2002)Kaverina, Krylyshkina, and Small]{kaverina02}
Kaverina, I., O.~Krylyshkina, and J.~Small, 2002.
\newblock Regulation of substrate adhesion dynamics during cell motility.
\newblock \emph{Int. J. Biochem. Cell. Biol.} 34:746 -- 761.

\bibitem[Parent and Devreotes(1999)]{parent99}
Parent, C.~A., and P.~N. Devreotes, 1999.
\newblock {A Cell's Sense of Direction}.
\newblock \emph{Science} 284:765--770.

\bibitem[Noegel and Schleicher(2000)]{noegel00}
Noegel, A., and M.~Schleicher, 2000.
\newblock {The actin cytoskeleton of Dictyostelium: a story told by mutants}.
\newblock \emph{J. Cell. Sci.} 113:759--766.

\bibitem[Kessin(2001)]{kessin01}
Kessin, R.~H., 2001.
\newblock {Dictyostelium : evolution, cell biology, and the development of
  multicellularity}.
\newblock Cambridge University Press, Cambridge, UK ; New York.

\bibitem[del {\'A}lamo et~al.(2007)del {\'A}lamo, Meili, Alonso-Latorre,
  Rodr{\'i}guez-Rodr{\'i}guez, Aliseda, Firtel, and Lasheras]{alamo07}
del {\'A}lamo, J.~C., R.~Meili, B.~Alonso-Latorre,
  J.~Rodr{\'i}guez-Rodr{\'i}guez, A.~Aliseda, R.~A. Firtel, and J.~C. Lasheras,
  2007.
\newblock {Spatio-temporal analysis of eukaryotic cell motility by improved
  force cytometry}.
\newblock \emph{Proc. Natl. Acad. Sci.} 104:13343--13348.

\bibitem[Lombardi et~al.(2007)Lombardi, Knecht, Dembo, and Lee]{lombardi07}
Lombardi, M.~L., D.~A. Knecht, M.~Dembo, and J.~Lee, 2007.
\newblock {Traction force microscopy in Dictyostelium reveals distinct roles
  for myosin II motor and actin-crosslinking activity in polarized cell
  movement}.
\newblock \emph{J. Cell. Sci.} 120:1624--1634.

\bibitem[Uchida and Yumura(2004)]{uchida04}
Uchida, K. S.~K., and S.~Yumura, 2004.
\newblock {Dynamics of novel feet of Dictyostelium cells during migration}.
\newblock \emph{J. Cell. Sci.} 117:1443--1455.

\bibitem[Iwadate and Yumura(2008)]{iwadate08}
Iwadate, Y., and S.~Yumura, 2008.
\newblock {Actin-based propulsive forces and myosin-II-based contractile forces
  in migrating Dictyostelium cells}.
\newblock \emph{J. Cell. Sci.} 121:1314--1324.

\bibitem[Simson et~al.(1998)Simson, Wallraff, Faix, Niew{\"o}hner, Gerisch, and
  Sackmann]{simson98}
Simson, R., E.~Wallraff, J.~Faix, J.~Niew{\"o}hner, G.~Gerisch, and
  E.~Sackmann, 1998.
\newblock Membrane Bending Modulus and Adhesion Energy of Wild-Type and Mutant
  Cells of Dictyostelium Lacking Talin or Cortexillins.
\newblock \emph{Biophys. J.} 74:514 -- 522.

\bibitem[D{\'e}cav{\'e} et~al.(2002{\natexlab{a}})D{\'e}cav{\'e}, Garrivier,
  Br{\'e}chet, Fourcade, and Bruckert]{decave02a}
D{\'e}cav{\'e}, E., D.~Garrivier, Y.~Br{\'e}chet, B.~Fourcade, and F.~Bruckert,
  2002.
\newblock Shear Flow-Induced Detachment Kinetics of Dictyostelium discoideum
  Cells from Solid Substrate.
\newblock \emph{Biophys. J.} 82:2383 -- 2395.

\bibitem[Jay et~al.(1995)Jay, Pham, Wong, and Elson]{jay95}
Jay, P.~Y., P.~A. Pham, S.~A. Wong, and E.~L. Elson, 1995.
\newblock {A mechanical function of myosin II in cell motility}.
\newblock \emph{J. Cell. Sci.} 108:387--393.

\bibitem[Duran et~al.(2009)Duran, Rahman, Colten, and Brazill]{duran09}
Duran, M.~B., A.~Rahman, M.~Colten, and D.~Brazill, 2009.
\newblock Dictyostelium discoideum Paxillin Regulates Actin-Based Processes.
\newblock \emph{Protist} 160:221 -- 232.

\bibitem[DiMilla et~al.(1991)DiMilla, Barbee, and Lauffenburger]{dimilla91}
DiMilla, P.~A., K.~Barbee, and D.~A. Lauffenburger, 1991.
\newblock Mathematical model for the effects of adhesion and mechanics on cell
  migration speed.
\newblock \emph{Biophys. J.} 60:15 -- 37.

\bibitem[Bottino and Fauci(1998)]{bottino98}
Bottino, D.~C., and L.~J. Fauci, 1998.
\newblock {A computational model of ameboid deformation and locomotion}.
\newblock \emph{Eur. Biophys. J.} 27:p532 --.

\bibitem[Gracheva and Othmer(2004)]{gracheva04}
Gracheva, M.~E., and H.~G. Othmer, 2004.
\newblock A continuum model of motility in ameboid cells.
\newblock \emph{Bull. Math. Biol.} 66:167 -- 193.

\bibitem[Larripa and Mogilner(2006)]{larripa06}
Larripa, K., and A.~Mogilner, 2006.
\newblock Transport of a 1D viscoelastic actin-myosin strip of gel as a model
  of a crawling cell.
\newblock \emph{Phys. A} 372:113 -- 123.

\bibitem[Palecek et~al.(1997)Palecek, Loftus, Ginsberg, Lauffenburger, and
  Horwitz]{palecek97}
Palecek, S.~P., J.~C. Loftus, M.~H. Ginsberg, D.~A. Lauffenburger, and A.~F.
  Horwitz, 1997.
\newblock Integrin-ligand binding properties govern cell migration speed
  through cell-substratum adhesiveness.
\newblock \emph{Nature} 385:537--540.

\bibitem[Bottino et~al.(2002)Bottino, Mogilner, Roberts, Stewart, and
  Oster]{bottino02}
Bottino, D., A.~Mogilner, T.~Roberts, M.~Stewart, and G.~Oster, 2002.
\newblock {How nematode sperm crawl}.
\newblock \emph{J. Cell. Sci.} 115:367--384.

\bibitem[Riveline et~al.(1998)Riveline, Ott, J{\"u}licher, Winkelmann, Cardoso,
  Lacap{\`e}re, Magn{\'u}d{\'o}ttir, Viovy, Gorre-Talini, and
  Prost]{riveline98}
Riveline, D., A.~Ott, F.~J{\"u}licher, D.~A. Winkelmann, O.~Cardoso, J.-J.
  Lacap{\`e}re, S.~Magn{\'u}d{\'o}ttir, J.~L. Viovy, L.~Gorre-Talini, and
  J.~Prost, 1998.
\newblock {Acting on actin: the electric motility assay}.
\newblock \emph{Eur. Biophys. J.} 27:p403 --.

\bibitem[H{\"a}nggi et~al.(1990)H{\"a}nggi, Talkner, and Borkovec]{haenggi90}
H{\"a}nggi, P., P.~Talkner, and M.~Borkovec, 1990.
\newblock Reaction-rate theory: fifty years after Kramers.
\newblock \emph{Rev. Mod. Phys.} 62:251--341.

\bibitem[Sabouri-Ghomi et~al.(2008)Sabouri-Ghomi, Wu, Hahn, and
  Danuser]{sabouri08}
Sabouri-Ghomi, M., Y.~Wu, K.~Hahn, and G.~Danuser, 2008.
\newblock Visualizing and quantifying adhesive signals.
\newblock \emph{Curr. Opin. Cell Biol.} 20:541 -- 550.

\bibitem[Huttenlocher et~al.(1997)Huttenlocher, Palecek, Lu, Zhang, Mellgren,
  Lauffenburger, Ginsberg, and Horwitz]{huttenlocher97}
Huttenlocher, A., S.~P. Palecek, Q.~Lu, W.~Zhang, R.~L. Mellgren, D.~A.
  Lauffenburger, M.~H. Ginsberg, and A.~F. Horwitz, 1997.
\newblock {Regulation of Cell Migration by the Calcium-dependent Protease
  Calpain}.
\newblock \emph{J. Biol. Chem.} 272:32719--32722.

\bibitem[Bell(1978)]{bell78}
Bell, G.~I., 1978.
\newblock {Models for the specific adhesion of cells to cells}.
\newblock \emph{Science} 200:618--627.

\bibitem[D{\'e}cav{\'e} et~al.(2002{\natexlab{b}})D{\'e}cav{\'e}, Garrivier,
  Br{\'e}chet, Bruckert, and Fourcade]{decave02b}
D{\'e}cav{\'e}, E., D.~Garrivier, Y.~Br{\'e}chet, F.~Bruckert, and B.~Fourcade,
  2002.
\newblock {Peeling Process in Living Cell Movement Under Shear Flow}.
\newblock \emph{Phys. Rev. Lett.} 89:108101.

\bibitem[Marshall et~al.(2006)Marshall, Sarangapani, Wu, Lawrence, McEver, and
  Zhu]{marshall06}
Marshall, B.~T., K.~K. Sarangapani, J.~Wu, M.~B. Lawrence, R.~P. McEver, and
  C.~Zhu, 2006.
\newblock Measuring Molecular Elasticity by Atomic Force Microscope Cantilever
  Fluctuations.
\newblock \emph{Biophys. J.} 90:681--692.

\bibitem[Schindl et~al.(1995)Schindl, Wallraff, Deubzer, Witke, Gerisch, and
  Sackmann]{schindl95}
Schindl, M., E.~Wallraff, B.~Deubzer, W.~Witke, G.~Gerisch, and E.~Sackmann,
  1995.
\newblock Cell-substrate interactions and locomotion of Dictyostelium wild-type
  and mutants defective in three cytoskeletal proteins: a study using
  quantitative reflection interference contrast microscopy.
\newblock \emph{Biophys. J.} 68:1177 -- 1190.

\bibitem[Weber et~al.(1995)Weber, Wallraff, Albrecht, and Gerisch]{weber95}
Weber, I., E.~Wallraff, R.~Albrecht, and G.~Gerisch, 1995.
\newblock {Motility and substratum adhesion of Dictyostelium wild-type and
  cytoskeletal mutant cells: a study by RICM/bright-field double-view image
  analysis}.
\newblock \emph{J. Cell. Sci.} 108:1519--1530.

\bibitem[Murphy et~al.(2001)Murphy, Rock, and Spudich]{murphy01}
Murphy, C.~T., R.~S. Rock, and J.~A. Spudich, 2001.
\newblock A myosin II mutation uncouples ATPase activity from motility and
  shortens step size.
\newblock \emph{Nat. Cell Biol.} 3:311--315.

\bibitem[Bukharova et~al.(2005)Bukharova, Bukahrova, Weijer, Bosgraaf, Dormann,
  van Haastert, and Weijer]{bukharova05}
Bukharova, T., T.~Bukahrova, G.~Weijer, L.~Bosgraaf, D.~Dormann, P.~J. van
  Haastert, and C.~J. Weijer, 2005.
\newblock Paxillin is required for cell-substrate adhesion, cell sorting and
  slug migration during Dictyostelium development.
\newblock \emph{J. Cell. Sci.} 118:4295--4310.

\end{thebibliography}
\end{document}